\documentclass[11pt]{article}
\usepackage{graphicx}
\usepackage{amsmath,amsthm,amssymb}
\begin{document}
\title{\textbf{\textsf{Generalized Holographic Dark Energy Model} }}
\author{ Mubasher Jamil\footnote{mjamil@camp.edu.pk},\ \ M. Umar
Farooq and Muneer Ahmad Rashid\footnote{muneerrshd@yahoo.com}
\\ \\
\small Center for Advanced Mathematics and Physics, National University of Sciences and Technology\\
\small E\&ME Campus, Peshawar Road, Rawalpindi, 46000, Pakistan \\
} \maketitle
\begin{abstract}
In this paper, the model of the holographic Chaplygin gas has been
extended to two general cases: first the case of a modified variable
Chaplygin gas and second the case of the viscous generalized
Chaplygin gas. The dynamics of the model is expressed by the use of
scalar fields and scalar potentials.
\end{abstract}
\textit{Keywords}: Cosmological constant; cosmic coincidence
problem; dark energy; holographic principle.

PACS numbers: 98.80.Cq, 98.80.-k, 98.80.Jk

\newpage
\large

\section{Introduction}

Astrophysical observations of type Ia supernovae \cite{perl,riess},
galaxy redshift surveys \cite{Fedeli}, cosmic background radiation
(CMB) data \cite{Keum,huang12,caldwell2}, large scale structure
\cite{mota22,daniel} and gravitational lensing surveys
\cite{Hoekstra} convincingly suggest that the observable universe is
undergoing accelerated expansion. The observations also suggest that
the transition from the earlier deceleration phase to the recent
acceleration phase is marginally recent $z\lesssim1$
\cite{caldwell}. The cause of this sudden transition and the source
of this expansion has not yet been identified categorically. It is
generally believed that some sort of `dark energy' is pervading the
universe. Consequently several questions arise: if dark energy
dominates the universe, then why did it remain dormant until
recently? Can it interact with other cosmic ingredients like matter
and radiation? What is its equation of state (EoS)? and also what is
the composition of this energy? The word `dark' itself implies that
our understanding about the nature of this energy is very modest,
despite substantial progress both in the theoretical and
observational fields. Several cosmological models have been proposed
in recent years to explain dark energy including those based on the
Chaplygin gas \cite{Compean}, scalar field models like quintessence
\cite{tomi,caldwell1}, k-essence \cite{thomas,picon} and phantom
energy \cite{jamil}, modified $f(R)$ gravity theories
\cite{nojiri,nojiri1} and variable constants approach \cite{jamil2},
to name a few. It might be possible that the accelerated expansion
of the universe is a manifestation of the inhomogeneity and
anisotropy of the space itself and that dark energy may not be
mandatory at all \cite{ltb}.

The problem of dark energy has also been addressed in the context of
the holographic principle. The principle says that the number of
degrees of freedom of a physical system should scale with its
bounding area rather than with its volume. It is motivated from an
observation that in quantum field theory, the ultraviolet cut-off
$\Lambda$ could be related to the infrared cut-off $L$ due to the
limit set by forming a black hole i.e. the total energy of a system
of size $L$ should not exceed the mass of a system-size black hole
\cite{cohen,myung}:
\begin{equation}
L^3\rho_\Lambda\leq M_p^2L,
\end{equation}
which yields
\begin{equation}
\rho_\Lambda=3c^2M_p^2L_\Lambda^{-2}.
\end{equation}
Here $c$ is a positive constant of the order unity and $3c^2$ is
attached for convenience, $M_p\equiv(8\pi G)^{-1/2}$ is the reduced
Planck mass and $L_\Lambda$ is the largest cut-off chosen to convert
into an equality. It has been shown that a
Friedmann-Robertson-Walker (FRW) universe filled with matter and
holographic dark energy (HDE) fits the supernova Ia data if the
parameter $c$ is taken to be $c=0.21$ or more generally $c<1$
\cite{zhang,zhang1}. We shall use the notation $\rho_\Lambda$ and
$\rho_{de}$ synonymously. Eq. (2) represents the energy density
corresponding to the HDE. Hence if the infrared cut-off is taken as
the current size of the universe i.e. $L_\Lambda=H^{-1}$ then the
holographic energy density is close to the density of dark energy
\cite{horava}. In \cite{setare1}, the author has developed the
correspondence between HDE and tachyons while in \cite{setare2}, the
connection between HDE and Gauss-Bonnet dark energy is established.
In \cite{wu}, the relationship between HDE and $f(R)$ theories is
developed, and also in \cite{gong} the analogy between HDE and
Brans$-$Dicke theory is proposed, while in \cite{granda}, a
correspondence of HDE with quintessence, tachyons and K-essence is
obtained. All these correspondences lead to accelerated expansion
solutions at late times. The HDE can also realize a quintom scenario
i.e. it evolves from a quintessence phase to the phantom phase
\cite{zhang2,li}. The holographic dark energy has been tested and
constrained by various observations, such as SNe Ia \cite{huang},
CMB \cite{shen}, X-ray gas mass fraction of galaxy clusters
\cite{chang} and the differential ages of passively evolving
galaxies \cite{yi}. The HDE also fairly alleviates some hard
cosmological problems like the cosmic age problem \cite{wei},the
cosmic coincidence problem \cite{karvan,cruz} and the fine tuning
problem \cite{feng}.

The holographic dark energy model was originally proposed by Nojiri
and Odintsov \cite{nojiri2}. Recently, the holographic dark energy
model with Chaplygin gas \cite{setare} and with modified Chaplygin
gas \cite{paul} have been investigated. We here extend their studies
to two EoSs involving a modified variable Chaplygin gas and viscous
dark energy in the context of the holographic principle, in the next
two sections. These EoSs belong to a general class of inhomogeneous
EoSs as discussed in \cite{nojiri3}.

\section{Holographic modified variable Chaplygin gas model}

We start by assuming the background to be spatially homogeneous and
isotropic FRW spacetime, specified by the line element
\begin{equation}
ds^2=-dt^2+a^2(t)\left[
\frac{dr^2}{1-kr^2}+r^2(d\theta^2+\sin^2\theta d\phi^2) \right].
\end{equation}
Here $a(t)$ is the dimensionless scale factor and $k$ is the
curvature parameter which takes the three possible values $+1,0,-1$
which correspond to spatially closed, flat and open spacetimes,
respectively. The corresponding Einstein field equation is given by
\begin{equation}
H^2+\frac{k}{a^2}=\frac{1}{3M_p^2}\left[\rho_{de}+\rho_m\right],
\end{equation}
where $M_p^2=(8\pi G)^{-1}$ is the modified Planck mass. The
dimensionless density parameters corresponding to matter, dark
energy and curvature are
\begin{equation}
\Omega_m=\frac{\rho_m}{\rho_{cr}}=\frac{\rho_m}{3H^2M_p^2},\ \
\Omega_{de}=\frac{\rho_{de}}{\rho_{cr}}=\frac{\rho_{de}}{3H^2M_p^2},\
\ \Omega_k=\frac{k}{(aH)^2}.
\end{equation}
Here $\rho_{cr}=3M_p^2H^2$ is the critical density. Note that from
(4) and (5), we can write $\Omega_{m}+\Omega_{de}=1+\Omega_k$. The
EoS representing the dark energy is the modified variable Chaplygin
gas (MVG) and is given by \cite{debnath,jamil4}
\begin{equation}
p_{de}=A\rho_{de}-\frac{B(a)}{\rho_{de}^\alpha},\ \ B(a)=B_oa^{-n}.
\end{equation}
Here $0\leq\alpha\leq1$, $0\leq A\leq1$, $B_o$ and $n$ are constant
parameters. The Chaplygin gas behaves like dust in the early
evolution of the universe and subsequently grows to an asymptotic
cosmological constant at late time when the universe is sufficiently
large. In the cosmological context, the Chaplygin gas was first
suggested as an alternative to quintessence \cite{kamenshchik}.
Later on, the Chaplygin gas state equation was extended to a
modified form by adding a barotropic term \cite{Benaoum,debnath2}.
Recent supernovae data also favor the two-fluid cosmological model
with Chaplygin gas and matter \cite{grigoris}. We assume the two
species i.e. matter and dark energy to be non-interacting, thus the
energy conservation equations are
\begin{equation}
\dot{\rho}_m+3H\rho_m=0,\ \ \dot{\rho}_{de}+3H(\rho_{de}+p_{de})=0.
\end{equation}
The evolution of the energy density of MVG is obtained by
substituting (6) in (7) to get
\begin{equation}
\rho_{de}=\left[\frac{3(1+\alpha)B_o}{[3(1+\alpha)(1+A)-n]}\frac{1}{a^n}-\frac{C}{a^{3(1+\alpha)(1+A)}}
\right]^{\frac{1}{1+\alpha}}.
\end{equation}
Here $C$ is a constant of integration.

Astrophysical observations suggest that the EoS parameter
$\omega_{de}$ is a dynamical variable which favors a
phantom-non-phantom transition in the recent past. This behavior of
dark energy is best explained with the help of a dynamically
evolving and minimally coupled scalar field \cite{zhang22}. Note
that this kind of scalar field formalism for dark energy is
motivated by the cosmological inflation models as well \cite{coles}.
Consider a scalar field $\phi$ with potential $V(\phi)$, related
with the energy density and the pressure of MVG as
\begin{equation}
\rho_\phi=\frac{1}{2}\dot{\phi}^2+V(\phi)=\left[\frac{3(1+\alpha)B_o}{[3(1+\alpha)(1+A)-n]}\frac{1}{a^n}-\frac{C}{a^{3(1+\alpha)(1+A)}}
\right]^{\frac{1}{1+\alpha}}.
\end{equation}
\begin{eqnarray}
p_\phi=\frac{1}{2}\dot{\phi}^2-V(\phi)&=&A\left[\frac{3(1+\alpha)B_o}{[3(1+\alpha)(1+A)-n]}\frac{1}{a^n}-\frac{C}{a^{3(1+\alpha)(1+A)}}
\right]^{\frac{1}{1+\alpha}}\nonumber\\&\;&-\frac{B_oa^{-n}}{\left[\frac{3(1+\alpha)B_o}{[3(1+\alpha)(1+A)-n]}\frac{1}{a^n}-\frac{C}{a^{3(1+\alpha)(1+A)}}
\right]^{\frac{\alpha}{1+\alpha}}}.
\end{eqnarray}
From the last two equations, the kinetic and the potential terms are
evaluated to be
\begin{eqnarray}
\dot{\phi}^2=\rho_\phi+p_\phi&=&(1+A)\left[\frac{3(1+\alpha)B_o}{[3(1+\alpha)(1+A)-n]}\frac{1}{a^n}-\frac{C}{a^{3(1+\alpha)(1+A)}}
\right]^{\frac{1}{1+\alpha}}\nonumber\\&\;&-\frac{B_oa^{-n}}{\left[\frac{3(1+\alpha)B_o}{[3(1+\alpha)(1+A)-n]}\frac{1}{a^n}-\frac{C}{a^{3(1+\alpha)(1+A)}}
\right]^{\frac{\alpha}{1+\alpha}}}.
\end{eqnarray}
\begin{eqnarray}
2V(\phi)=\rho_\phi-p_\phi&=&(1-A)\left[\frac{3(1+\alpha)B_o}{[3(1+\alpha)(1+A)-n]}\frac{1}{a^n}-\frac{C}{a^{3(1+\alpha)(1+A)}}
\right]^{\frac{1}{1+\alpha}}\nonumber\\&\;&+\frac{B_oa^{-n}}{\left[\frac{3(1+\alpha)B_o}{[3(1+\alpha)(1+A)-n]}\frac{1}{a^n}-\frac{C}{a^{3(1+\alpha)(1+A)}}
\right]^{\frac{\alpha}{1+\alpha}}}.
\end{eqnarray}
The size of the future event horizon $R_h$ is defined as
\begin{equation}
R_h(t)=a(t)\int\limits_t^\infty\frac{dt^\prime}{a(t^\prime)}=a(t)\int\limits_0^{r_1}\frac{dr}{\sqrt{1-kr^2}}.
\end{equation}
We shall use the identity
\begin{equation}\int\limits_0^{r_1}\frac{dr}{\sqrt{1-kr^2}}=\frac{1}{\sqrt{|k|}}\text{sinn}^{-1}(\sqrt{|k|}r_1)=
\begin{cases} \text{sin}^{-1}r_1  & \, \,k=+1,\\
             r_1 & \, \,  k=0,\\
             \text{sinh}^{-1}r_1 & \, \,k=-1.\\
\end{cases}\end{equation}
Also $L$ is defined as
\begin{equation}
L=ar(t).
\end{equation}
It is suggested in \cite{li} that $r(t)=R_h(t)$ as the infrared
cut-off. From Eqs. (3), (13) and (15), we can write
\begin{equation}
L=a(t)\frac{\text{sinn}[\sqrt{|k|}R_h(t)/a(t)]}{\sqrt{|k|}}.
\end{equation}
Here $\text{sinn}y=\text{sin}y,y,\text{sin}^{-1}y$ for $k=+1,0,-1$
respectively, where $y=\sqrt{|k|}R_h(t)/a(t)$. Combining the usual
definitions of $\Omega_{de}$ and $\rho_{cr}$ yields
\begin{equation}
HL=\frac{c}{\sqrt{\Omega_{de}}}.
\end{equation}
Differentiating Eq. (17) with respect to $t$ and using (16), we get
\begin{equation}
\dot{L}=\frac{c}{\sqrt{\Omega_{de}}}-\frac{1}{\sqrt{|k|}}\text{cosn}(\sqrt{|k|}R_h/a).
\end{equation}
Here $\text{cosn}y=\text{cos}y,y,\text{cos}^{-1}y$ for $k=+1,0,-1$
respectively, where $y=\sqrt{|k|}R_h(t)/a(t)$. The EoS parameter for
dark energy is given by
\begin{equation}
\omega_{de}=-\frac{1}{3}-\frac{2\sqrt{\Omega_{de}}}{3c}\frac{1}{\sqrt{|k|}}\text{cosn}(\sqrt{|k|}R_h/a).
\end{equation}
Invoking the correspondence between Eqs. (2) and (8), we obtain
\begin{equation}
C=a^{3(1+\alpha)(1+A)}\left[ (3c^2M_p^2L^{-2})^{1+\alpha} -
\frac{3(1+\alpha)B_o}{[3(1+\alpha)(1+A)-n]}\frac{1}{a^n}\right].
\end{equation}
The constant parameter $B_o$ is determined to be
\begin{eqnarray}
B_o&=&a^n\rho_{de}^{1+\alpha}(A-\omega_{de}),\nonumber\\
&=&a^n(3c^2M_p^2L^{-2})^{1+\alpha}\left[ A+
\frac{1}{3}+\frac{2\sqrt{\Omega_{de}}}{3c}\frac{1}{\sqrt{|k|}}\text{cosn}(\sqrt{|k|}R_h/a)\right].
\end{eqnarray}
Using Eq. (21) in (20), we get
\begin{equation}
C=a^{3(1+\alpha)(1+A)}(3c^2M_p^2L^{-2})^{1+\alpha}\left[
1-\frac{3(1+\alpha)}{[(1+\alpha)(1+A)-n]} \left\{  A+
\frac{1}{3}+\frac{2\sqrt{\Omega_{de}}}{3c}\frac{1}{\sqrt{|k|}}\text{cosn}(\sqrt{|k|}R_h/a)
\right\}\right].
\end{equation}
The EoS parameter $\omega_{de}$ is
\begin{equation}
\omega_{de}=\frac{p_{de}}{\rho_{de}}=\frac{1}{\rho_{de}}\left[
A\rho_{de}-\frac{B_oa^{-n}}{\rho_{de}^\alpha}
\right]=A-\frac{B_oa^{-n}}{\rho_{de}^{1+\alpha}}.
\end{equation}
From Eq. (11), the kinetic term is re-written to be
\begin{equation}
\dot{\phi}^2=2c^2M^2_pL^{-2}\left[
1-\frac{\sqrt{\Omega_{de}}}{3c}\frac{1}{\sqrt{|k|}}\text{cosn}(\sqrt{|k|}R_h/a)
\right],
\end{equation}
\begin{equation}
\dot{\phi}=\sqrt{(2c^2M^2_pL^{-2})\left[
1-\frac{\sqrt{\Omega_{de}}}{3c}\frac{1}{\sqrt{|k|}}\text{cosn}(\sqrt{|k|}R_h/a)
\right]}.
\end{equation}
From Eq. (12), the potential term becomes
\begin{eqnarray}
2V(\phi)&=&
\frac{\rho_{de}^{1+\alpha}(1-A)+B_oa^{-n}}{\rho_{de}^\alpha},\nonumber\\
&=&3c^2M_p^2L^{-2}\left[
\frac{4}{3}+\frac{2}{3}\frac{\sqrt{\Omega_{de}}}{3c}\frac{1}{\sqrt{|k|}}\text{cosn}(\sqrt{|k|}R_h/a)
\right].
\end{eqnarray}
Hence the potential term turns out to be the same as that for the
holographic Chaplygin gas \cite{setare}. Using the relation with
$x=\ln a$, we have
\begin{equation}
\dot{\phi}=\phi^\prime H,
\end{equation}
and we obtain
\begin{equation}
\phi^\prime=M_p\sqrt{2\Omega_{de}\left[
1-\frac{\sqrt{\Omega_{de}}}{3c}\frac{1}{\sqrt{|k|}}\text{cosn}(\sqrt{|k|}R_h/a)
\right]}.
\end{equation}
After integration, we get
\begin{equation}
\phi(a)-\phi(a_o)=M_p\int\limits_o^{\ln a}\sqrt{2\Omega_{de}\left[
1-\frac{\sqrt{\Omega_{de}}}{3c}\frac{1}{\sqrt{|k|}}\text{cosn}(\sqrt{|k|}R_h/a)
\right]}dx.
\end{equation}

\section{Holographic viscous Chaplygin gas model}

In this section, we shall consider dark energy with non-zero bulk
viscosity. The role of viscosity has been widely discussed and is a
promising candidate to explain several cosmological puzzles,
especially dark energy. Notably, the viscous dark energy can explain
the high photon to baryon ratio \cite{night}, and it can lead to an
inflationary scenario in the early phase of the universe
\cite{waga}. The coefficient of viscosity should decrease as the
universe expands; moreover, its presence can explain the current
accelerated expansion \cite{ren,singh,fabris}. It has been pointed
out that viscous dark energy can drive expansion so rapid that it
may result in a catastrophic big rip singularity \cite{cataldo}. A
cosmological model with bulk viscosity also rules out the
possibility of a big bang singularity, which is as yet unexplained
\cite{belinski}. This model is also consistent with astrophysical
observations at the lower redshifts, and a viscous cosmic model
favors a standard cold dark matter model with cosmological constant
($\Lambda$CDM) in the later cosmic evolution \cite{hu}. The model
also presents the scenario of phantom crossing (or phantom divide)
i.e. the transition of parameter $\omega_{de}>-1$ to
$\omega_{de}<-1$ \cite{cruz1}. We consider the viscous dark energy
with the EoS \cite{beesham,jamil3}
\begin{equation}
p_{eff}=p_{de}+\Pi.
\end{equation}
Here $p_{de}$ is the barotropic pressure (which depends only on the
energy density of the fluid) while $\Pi=-\xi(\rho_{de})u^\mu_{;\mu}$
is the viscous pressure, $u^\mu$ is the four-velocity vector and
$\xi$ is the coefficient of bulk viscosity \cite{eckart}. In the FRW
model, the viscous pressure takes the form $\Pi=-3H\xi$
\cite{zimdahl}. In the cosmological context, the barotropic EoS can
be chosen to be the Chaplygin gas \cite{pun}. For the purpose of
generality, we specify the $p_{de}$ by the generalized Chaplygin
gas, so that (30) takes the form
\begin{equation}
p_{eff}=\frac{\chi}{\rho_{de}^\alpha}-3H\xi(\rho_{de}).
\end{equation}
In general, the viscosity coefficient can be of power-law form, i.e.
$\xi\sim\rho^n$ for $n\geq0$ and hence it yields a power-law
expansion of the scale factor \cite{barrow}. We also assume the
parametric form $\xi(\rho_{de})=\nu\rho_{de}^{1/2}$, where $\nu$ is
a constant parameter. Hence Eq. (31) becomes
\begin{equation}
p_{eff}=\frac{\chi}{\rho_{de}^\alpha}-3\nu H\rho_{de}^{1/2}.
\end{equation}
The EoS parameter gives
\begin{equation}
\omega_{de}=\frac{p_{eff}}{\rho_{de}}=\frac{\chi}{\rho_{de}^{1+\alpha}}-3\nu
H\rho_{de}^{-1/2},
\end{equation}
or we can write
\begin{equation}
\chi=\rho_{de}^{1+\alpha}[\omega_{de}+3\nu H\rho_{de}^{-1/2}].
\end{equation}
The corresponding energy conservation equation is
\begin{equation}
\rho_{de}=\left[
\frac{Da^{-3(1-\nu\gamma)(1+\alpha)}-\chi}{1-\nu\gamma}
\right]^{\frac{1}{1+\alpha}}.
\end{equation}
Here $D$ is a constant of integration. The parameters used are
\begin{equation}
\gamma=M_p^{-1}\sqrt{1-r_m},\ \ r_m\equiv\rho_{m}/\rho_{de}.
\end{equation}
Using Eqs. (2) and (19) in (35), we obtain
\begin{eqnarray}
D&=&a^{3(1+\alpha)(1-\nu\gamma)}[\rho_{de}^{1+\alpha}(1-\nu\gamma)+\chi]\\
&=&a^{3(1+\alpha)(1-\nu\gamma)}(3c^2M_p^2L^{-2})^{1+\alpha}\left[\frac{2}{3}-\nu\gamma-\frac{2}{3}\frac{\sqrt{\Omega_{de}}}{c}\text{cosn}(\sqrt{|k|}R_h/a)+3\nu
H(3c^2M_p^2L^{-2})^{-1/2}\right]\nonumber.
\end{eqnarray}
The kinetic term becomes
\begin{eqnarray}
\dot{\phi}^2&=&\rho_\phi+p_\phi,\nonumber\\
&=&\frac{1}{\rho_{de}^\alpha}[\rho_{de}^{1+\alpha}+\chi-3\nu H\rho_{de}^{\alpha+\frac{1}{2}}],\nonumber\\
&=&(3c^2M_p^2L^{-2})\left\{
\frac{2}{3}-\frac{2}{3}\frac{\sqrt{\Omega_{de}}}{c}\frac{1}{\sqrt{|k|}}\text{cos}(\sqrt{|k|R_h/a})+\frac{3\nu
L}{c\sqrt{\rho_{cr}}}\right\}\nonumber\\&\;&- \frac{3\nu
c\sqrt{\rho_{cr}}}{L}.
\end{eqnarray}
Also, the potential term becomes
\begin{eqnarray}
2V(\phi)&=&(\rho_{de}-p_{eff}),\nonumber\\
&=&\frac{1}{\rho_{de}^\alpha}(\rho_{de}^{1+\alpha}-\chi+3\nu H\rho_{de}^{\alpha+1/2}).\nonumber\\
&=&(3c^2M_p^2L^{-2})\left\{
\frac{4}{3}+\frac{2}{3}\frac{\sqrt{\Omega_{de}}}{c}\frac{1}{\sqrt{|k|}}\text{cos}(\sqrt{|k|R_h/a})-\frac{3\nu
L}{c\sqrt{\rho_{cr}}} \right\}\nonumber\\&\;&+\frac{3\nu
c\sqrt{\rho_{cr}}}{L}.
\end{eqnarray}
From (38), we have
\begin{equation}
\dot{\phi}=HM_p\left[3\Omega_{de}\left\{
\frac{2}{3}-\frac{2}{3}\frac{\sqrt{\Omega_{de}}}{c}\frac{1}{\sqrt{|k|}}\text{cos}(\sqrt{|k|R_h/a})+\frac{3\nu
L}{c\sqrt{\rho_{cr}}} \right\}- \frac{3\nu
c\sqrt{\rho_{cr}}}{L}\right]^{1/2}.
\end{equation}
Making use of (40) in (27) gives
\begin{equation}
\phi^\prime=M_p\left[3\Omega_{de}\left\{
\frac{2}{3}-\frac{2}{3}\frac{\sqrt{\Omega_{de}}}{c}\frac{1}{\sqrt{|k|}}\text{cos}(\sqrt{|k|R_h/a})+\frac{3\nu
L}{c\sqrt{\rho_{cr}}}\right\}- \frac{3\nu
c\sqrt{\rho_{cr}}}{L}\right]^{1/2}.
\end{equation}
On integration, we obtain
\begin{eqnarray}
\phi(a)-\phi(a_o)&=&\int\limits_o^{\ln
a}M_p\left[3\Omega_{de}\left\{
\frac{2}{3}-\frac{2}{3}\frac{\sqrt{\Omega_{de}}}{c}\frac{1}{\sqrt{|k|}}\text{cos}(\sqrt{|k|R_h/a})+\frac{3\nu
L}{c\sqrt{\rho_{cr}}} \right\}- \frac{3\nu
c\sqrt{\rho_{cr}}}{L}\right]^{1/2}.
\end{eqnarray}

\section{Conclusion and discussion}

The holographic dark energy presents the dynamical nature of the
vacuum energy. This dynamical nature is manifested through the
holographic parameter $c$ which by varying gives an evolving dark
energy. For instance, if $c\geq1$ gives the quintessence where its
state equation parameter lies in the range $-1\leq\omega\leq-1/3$,
while $c=1$ yields the cosmological constant phase and $c<-1$ gives
the phantom energy dominated universe. Thus the whole range of $c$
provides a quintom (quintessence to phantom) like model.

As discussed before, several authors have established the connection
between the holographic dark energy and various theories of gravity.
These correspondences are motivated to demystify the origin of dark
energy and the evolution of the universe. In this context, we have
presented the link between the holographic dark energy and the
modified variable Chaplygin gas and the viscous generalized
Chaplygin gas. The Chaplygin gas has been extensively used in recent
literature on cosmology due to the fact that its predictions are
consistent with the observational results. Moreover, it gives a
unified picture of dark energy and dark matter, which helps in
building and analyzing new cosmological models. For specific choices
of parameters, our results in both models reduced to those discussed
in \cite{setare} and \cite{paul}, thus our model is an extension of
these previous studies. Finally, the dynamics of dark energy in our
model is described by scalar fields with scalar potentials.

\subsubsection*{Acknowledgments}
We would like to thank M.R. Setare, X. Zhang and S. Odintsov for
pointing out to us useful references during this work. We would also
like to thank the anonymous referee for useful criticism on this
work.

\end{document}